\documentclass[12pt]{iopart}
\usepackage{graphicx}

\begin{document}
\def \ee {\varepsilon}
\thispagestyle{empty}
\title[]{
New approach to the
thermal Casimir force between real metals
}

\author{
V.~M.~Mostepanenko${}^{1,2}$ and
B.~Geyer${}^{1}$
}

\address{${}^1$Center of Theoretical Studies and Institute for Theoretical
Physics, Leipzig University,
D-04009, Leipzig, Germany}

\address{${}^2$
Noncommercial Partnership  ``Scientific Instruments'',
Tverskaya St. 11, Moscow, 103905, Russia}

\begin{abstract}
The new approach to the theoretical description of the thermal
Casimir force between real metals is presented. It uses the
plasma-like dielectric permittivity that takes into account the
interband transitions of core electrons. This permittivity
precisely satisfies the Kramers-Kronig relations.
The respective Casimir entropy is positive and vanishes
at zero temperature in accordance with the Nernst heat theorem.
The physical reasons why the Drude dielectric function, when
substituted in the Lifshitz formula, is inconsistent with
electrodynamics are elucidated. The proposed approach is the
single one consistent with all measurements of the Casimir
force performed up to date.
The application of this approach to metal-type semiconductors
is considered.
\end{abstract}

\section{Introduction}
The Casimir force \cite{1} acts between two parallel electrically neutral
metal plates in vacuum. This effect is entirely quantum. There is
no such force in classical electrodynamics. In accordance to quantum
field theory, there are zero-point oscillations of the electromagnetic
field in the vacuum state. The Casimir effect arises due to the boundary
conditions imposed on the electromagnetic field on metal surfaces.
The spectra of zero-point oscillations in the presence and in the
absence of plates are different. Casimir was the first who found the
finite difference between respective infinite vacuum energies.
The negative derivative of this difference with respect to the separation
between the plates is just what is referred to as the
{\it Casimir force}.

In his famous paper \cite{1} Casimir considered ideal metal
plates at zero temperature. Modern progress in measurements of the
Casimir force (see early stages in review \cite{2} and later
experiments \cite{3,4,5,5a,5b,6,7}) demand consideration of realistic
plates of finite conductivity at nonzero temperature. This is also
of much importance for the applications of the Casimir effect in
nanotechnology \cite{8,9}. The basic theory of both the van der Waals
and Casimir force taking into account the effects of finite
conductivity and nonzero temperature was developed by Lifshitz \cite{10}.
It describes material properties by means of the frequency dependent
dielectric permittivity.
First applications of this theory at nonzero temperature using the
Drude \cite{11} and plasma \cite{12,13} models for the dielectric
permittivity have led, however, to contradictory results.
In particular, thermal Casimir force if calculated using the Drude model
was found to be in qualitative disagreement with the case of
ideal metals. The large thermal correction arising at short separation
distances within this approach was excluded experimentally \cite{5,5a,5b}.
In addition, the Casimir entropy calculated using the Drude model
violates the third law of thermodynamics (the Nernst heat theorem) for
perfect crystal lattices with no impurities \cite{14,15}.
As to the nondissipative plasma model, it leads to the thermal Casimir
force in qualitative agreement with the case of ideal metals and
satisfies the Nernst theorem. It was found to be consistent with
the data of relatively large separation experiments on the measurement
of the Casimir force \cite{5,5a,5b}. However, as is demonstrated
below in Section 2, it is in contradiction with the results of short
separation experiment \cite{16}.
This can be explained by the fact that the plasma model completely
disregards interband transitions of core electrons.
Another approach to the thermal Casimir force
is based on the use of the Leontovich surface impedance
instead of dielectric permittivity \cite{17}.
This approach is consistent with thermodynamics and large separation
experiments, but it is not applicable at short separations.

In this paper we present and further elaborate a new theoretical
approach to the thermal Casimir force based on the use of generalized
plasma-like dielectric permittivity \cite{18}.
We demonstrate that this approach is consistent with all available
experimental results. Physical reasons why the Drude dielectric function
is not compatible with the Lifshitz theory in the case of finite
plates \cite{19} are discussed. We demonstrate that results obtained
in \cite{19} have far reaching consequences not only for metals but
also for semiconductors of metallic type with sufficiently high
dopant concentration.

In Section 2, on the basis of fundamentals of statistical physics
and all available experimental data we explain why the new approach
to the thermal Casimir force is much needed. Section 3 explains
the effect of finite plates, i.e., that the Drude dielectric
function is not applicable when the front of an incident wave has
much larger extension than the size of the plate. Section 4 contains
the formulation of the generalized plasma-like dielectric permittivity.
The results of a recent 6-oscillator fit to its parameters (see
Ref.~\cite{5b}) using the tabulated optical data for Au \cite{20}
are also presented. In Section 5 we compare the generalized
Kramers-Kronig relations valid for the plasma-like permittivity with
the standard ones valid for dielectrics and for Drude metals.
Section 6 briefly presents the thermodynamic test for the generalized
plasma-like permittivity. In Section 7 we apply the developed approach
to the case of semiconductors with relatively high concentration of
charge carriers. Section 8 contains our conclusions and discussion.

\section{Why a new approach is needed?}

We start with the Lifshitz formula for the free energy of the van der
Waals and Casimir interaction between the two parallel metallic plates
of thickness $d$ at a separation distance $a$ at temperature $T$ in
thermal equilibrium
\begin{eqnarray}
&&
{\cal F}(a,T)=\frac{k_BT}{2\pi}\sum\limits_{l=0}^{\infty}
\left(1-\frac{1}{2}\delta_{0l}\right)
\int_{0}^{\infty}k_{\bot}\,dk_{\bot}
\nonumber \\
&&\phantom{aaa}
\times\left\{\ln\left[1-r_{\rm TM}^2({\rm i}\xi_l,k_{\bot})
e^{-2aq_l}\right]+
\ln\left[1-r_{\rm TE}^2({\rm i}\xi_l,k_{\bot})
e^{-2aq_l}\right]\right\}.
\label{eq1}
\end{eqnarray}
\noindent
Here
$k_B$ is the Boltzmann constant and $\xi_l=2\pi k_B Tl/\hbar$
with $l=0,\,1,\,2,\,\ldots$
are the Matsubara frequencies
($k_{\bot}=|\mbox{\boldmath$k$}_{\bot}|$ is the projection
of a wave vector on the plane of the plates).
The reflection coefficients for the two independent
polarizations of the electromagnetic field (transverse
magnetic and transverse electric) are given by
\begin{eqnarray}
&&
r_{\rm TM}({\rm i}\xi_l,k_{\bot})=
\frac{\varepsilon_l^2q_l^2-k_l^2}{\varepsilon_l^2q_l^2+
k_l^2+2q_lk_l\varepsilon_l\coth(k_ld)},
\nonumber \\
&&
r_{\rm TE}({\rm i}\xi_l,k_{\bot})=
\frac{k_l^2-q_l^2}{q_l^2+k_l^2+2q_lk_l\coth(k_ld)},
\label{eq2}
\end{eqnarray}
\noindent
where
\begin{equation}
q_l=\sqrt{k_{\bot}^2+\frac{\xi_l^2}{c^2}}, \quad
k_l=\sqrt{k_{\bot}^2+\varepsilon_l\frac{\xi_l^2}{c^2}},
\quad \varepsilon_l=\varepsilon({\rm i}\xi_l)
\label{eq3}
\end{equation}
\noindent
and $\varepsilon(\omega)$ is the dielectric permittivity
of the plate material.

As was already discussed in the Introduction, in the framework of the
Lifshitz theory the calculation results strongly depend on the model
of a metal used. The source of discrepances are different
contributions from zero frequency [the term with $l=0$ in Eq.~(\ref{eq1})].
For ideal metal plates it holds $|\varepsilon|=\infty$ at any
frequencies, including $\xi_0=0$, and from (\ref{eq2}) one obtains
\begin{equation}
r_{\rm TM}({\rm i}\xi_l,k_{\bot})=r_{\rm TE}({\rm i}\xi_l,k_{\bot})=1.
\label{eq4}
\end{equation}
\noindent
For metals with $\varepsilon\sim 1/\xi$ when $\xi\to 0$ (this includes
but not reduces to the Drude model) from (\ref{eq2}) it follows
\cite{11,21}
\begin{equation}
r_{\rm TM}(0,k_{\bot})=1,
\qquad r_{\rm TE}(0,k_{\bot})=0.
\label{eq5}
\end{equation}
\noindent
For metals with $\varepsilon\sim \omega_p^2/\xi^2$ when $\xi\to 0$ ,
where  $\omega_p$ is the plasma frequency,
(\ref{eq2}) leads to a qualitatively different result for the transverse
electric reflection coefficient \cite{12,13},
\begin{eqnarray}
&&
r_{\rm TM}(0,k_{\bot})=1,
\label{eq6} \\
&&
r_{\rm TE}(0,k_{\bot})=\frac{\omega_p^2}{\omega_p^2+2k_{\bot}^2c^2+
2k_{\bot}c\sqrt{k_{\bot}^2c^2+\omega_p^2}
\coth\Bigl(\frac{d}{c}\sqrt{k_{\bot}^2c^2+\omega_p^2}\Bigr)}.
\nonumber
\end{eqnarray}
\noindent
As is seen from the comparison of (\ref{eq4}) and (\ref{eq5}), there is
a qualitative disagreement in the values of $r_{\rm TE}(0,k_{\bot})$.
This results in hundreds times larger thermal corrections at short
separation if one uses (\ref{eq5}) instead of (\ref{eq4}).
At large separations the magnitudes of the Casimir free energy and
pressure obtained by using (\ref{eq5}) are one half of those when
using (\ref{eq4}). At the same time all results obtained from
(\ref{eq4}) and (\ref{eq6}) are in qualitative agreement. This is
guaranteed by the fact that (\ref{eq6}) smoothly approaches (\ref{eq4})
when $\omega_p\to\infty$.
\begin{figure*}[t]
\vspace*{-12.4cm}
\includegraphics{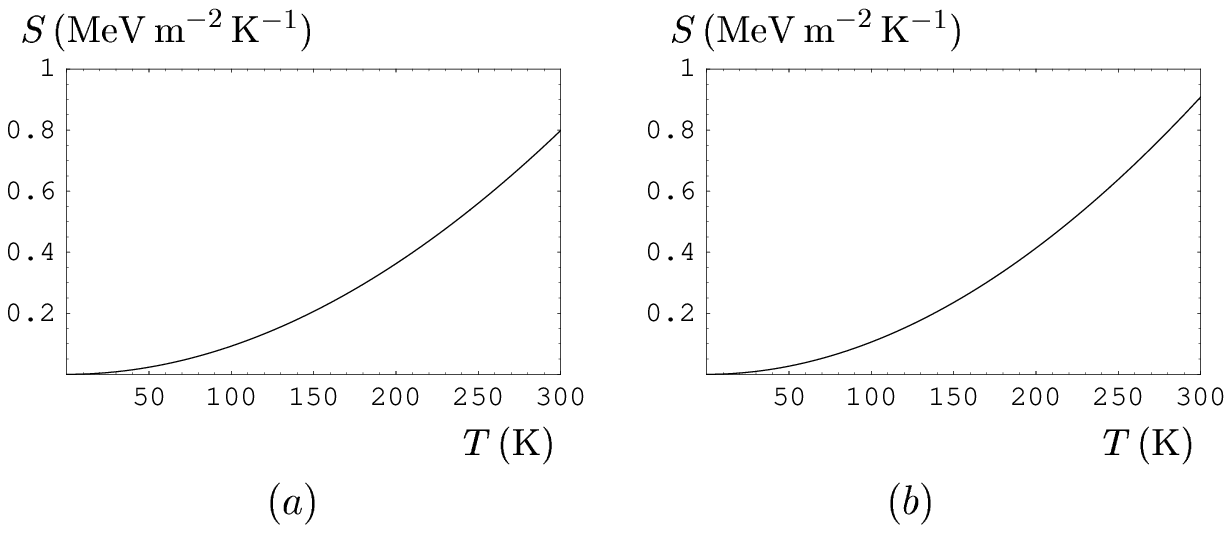}
\vspace*{-13.cm}
\caption{
The Casimir entropy for two plates made of an ideal metal (a) and
metal described by the plasma model with $\omega_p=9.0\,$eV (b)
at a separation 300\,nm.
}
\end{figure*}

The crucial question for any model used is its consistency with the
fundamental physical principles. In our case the considered models of
dielectric permittivity can be tested thermodynamically by the
behavior of the Casimir entropy,
\begin{equation}
S(a,T)=-\frac{\partial{\cal F}(a,T)}{\partial T},
\label{eq7}
\end{equation}
\noindent
at low temperatures. In figure 1(a) we plot the Casimir entropy
as a function of temperature for ideal metals \cite{22,23}
and in figure 1(b) for metals with
$\varepsilon\approx\omega_p^2/\xi^2$ when $\xi\to 0$ \cite{14,15}.
In both cases it holds $S(a,T)\geq 0$ and $S(a,T)\to 0$ when
$T$ vanishes. This means that the Nernst heat theorem is satisfied.

Quite different situation holds for metals with $\varepsilon\sim 1/\xi$
when $\xi\to 0$. The Casimir entropy for
$\varepsilon=1+\omega_p^2/[\xi(\xi+\gamma)]$, where $\gamma$ is the
relaxation parameter, is plotted in figure 2(a) \cite{14}. As is seen in this
figure,
the entropy becomes negative at $T$ of about several hundred K and
remains negative
\begin{equation}
S(a,0)=-\frac{k_B\zeta(3)}{16\pi a^2}\left[1-4\frac{c}{\omega_pa}+
12\Bigl(\frac{c}{\omega_pa}\Bigr)^2-\cdots\right]<0
\label{eq8}
\end{equation}
\noindent
at $T=0$ [here $\zeta(z)$ is the Riemann zeta function]. Thus, the Nernst
heat theorem is violated, suggesting that the model used is inapplicable.
Figure 2(a) is plotted for perfect crystal lattice with no impurities
when $\gamma\to 0$ with $T\to 0$. In \cite{24,24a} it was argued that
the presence of impurities can remedy this situation. However,
in \cite{24a} the dependence of the relaxation parameter on the temperature
was not taken into account. As a result, the coefficients of the
obtained asymptotic expressions were determined incorrectly up to
factors of several orders of magnitude \cite{25}.
For typical realistic concentrations of impurities the behavior
of the entropy as a function of temperature remains the same,
as in figure 2(a), up to as low temperatures as $10^{-3}-10^{-4}\,$K.
At lower temperatures, however, the Casimir entropy depends on $T$ in a
different way, as is shown in figure 2(b) for a typical residual
resistivity equal to $10^{-4}$ of the resistivity at room temperature.
Although from figure 2(b) it is seen that, at least formally, the
Nernst heat theorem is preserved for lattices with impurities, this
does not solve the contradiction between the models with
$\varepsilon\sim 1/\xi$ and thermodynamics. The point is that,
according to quantum statistical physics, the Nernst heat theorem
must be valid for perfect crystal lattice which has a nondegenerate
dynamic state of lowest energy. Thus, any model that violates this
rule is thermodynamically inacceptable.

Another crucial question for any model is consistency with experiment.
As is shown in \cite{5,5a,5b}, both the plasma model and the
impedance approach are consistent with the results of most precise
measurements of the Casimir pressure at separations $a\geq 160\,$nm.
The same measurement results are, however, inconsistent with the
Drude model approach. In figure 3(a) we plot the differences between the
theoretical Casimir pressures in the configuration of two parallel
plates calculated using the Drude model and tabulated optical data
and mean experimental pressures as a function of separation \cite{5b}.
It is seen that all differences are outside of the 95\% confidence
intervals whose boundaries generate a solid line. Within a wide
separation region from 210 to 620\,nm they are also outside of the
99.9\% confidence intervals indicated by the dashed line. Thus, the Drude
model approach is experimentally excluded.
\begin{figure*}[t]
\vspace*{-12.4cm}
\includegraphics{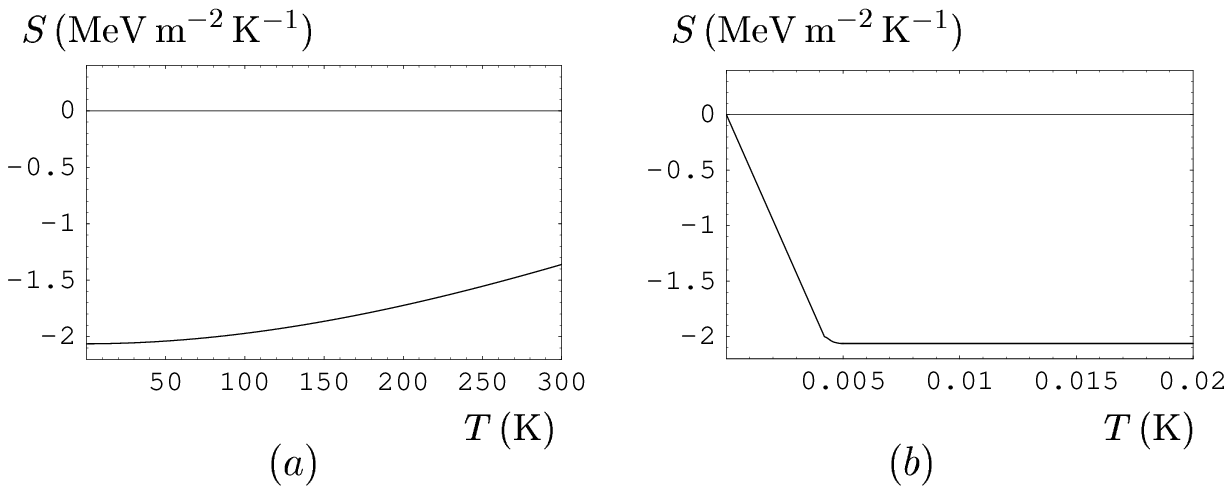}
\vspace*{-13.cm}
\caption{
The Casimir entropy for two plates made of Drude metal
with perfect crystal lattice (a) and of Drude
metal with impurities (b)
at a separation $1\,\mu$m ($\gamma=0.035\,$eV).
}
\end{figure*}
Note that in theoretical computations in figure 3($a$) the
tabulated optical data were extrapolated to low frequencies
by the Drude model with the plasma frequency
$\omega_p=8.9\,$eV and the relaxation parameter
$\gamma=0.0357\,$eV. Importantly, variations of $\gamma$
practically do not influence the magnitudes of the Casimir
pressure. For example, a descrease of $\gamma$ until 0.02\,eV
would lead to only 0.29\% increase of $|P_D^{\rm th}|$ at
$a=200\,$nm and to 0.26\% increase at $a=650\,$nm.
This is because the value of $\gamma$ does not influence the
zero-frequency term of the Lifshitz formula. As a result,
the width of separation intervals, where the Drude model
approach is excluded, practically does not depend on the
value of $\gamma$. If the smaller values of $\omega_p$ are
used (as suggested in \cite{28a}), the Drude model approach
is excluded at 99.9\% confidence level within even
wider separation interval.

The plasma model approach, although it agrees with the measurements
of \cite{5,5a,5b}, also cannot be considered as a universal.
In figure 3(b) we plot the differences of the theoretical Casimir
forces between a plate and a sphere
which are calculated using the plasma model
and mean experimental Casimir forces \cite{16} versus separation.
It is seen that at separations below 80\,nm the plasma model approach
is excluded by the experimental data. Bearing in mind that at so
short separations (below the plasma wavelength) the impedance
approach is not applicable, it may be concluded that until recently
there was no theoretical approach to the thermal Casimir force
consistent with both long-separation and short-separation
experiments. Such an approach based on the generalized plasma-like
permittivity was first proposed in \cite{18,19} and used in \cite{5b}.
Below we discuss the main points of this approach and apply it to
semiconductors of metallic type.
\begin{figure*}[t]
\vspace*{-14.4cm}
\includegraphics{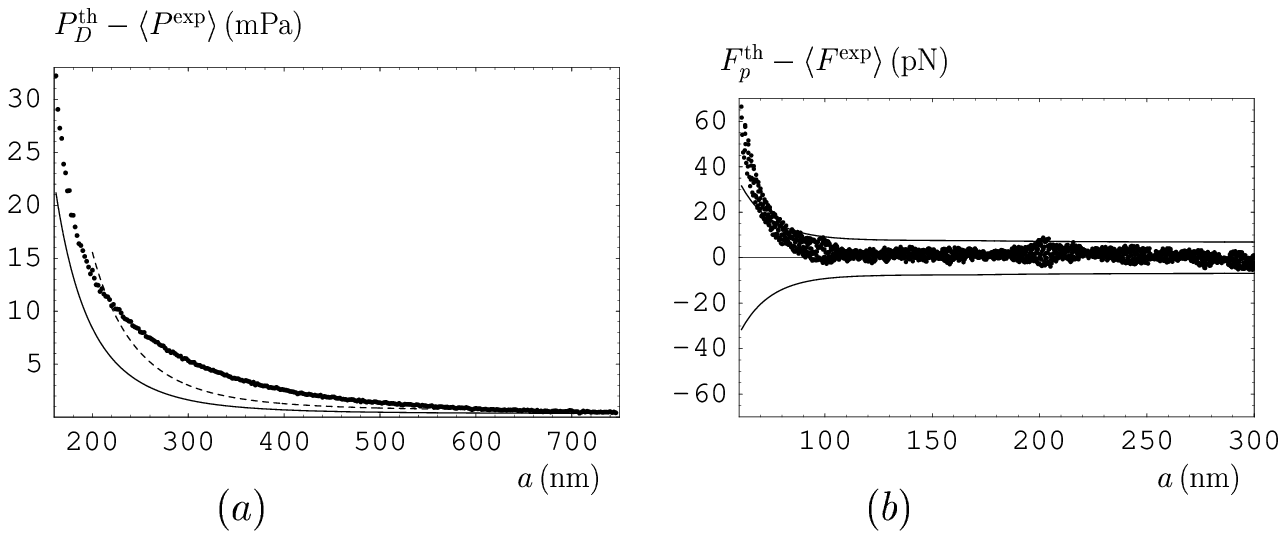}
\vspace*{-11.cm}
\caption{
Differences between the theoretical and mean experimental Casimir
pressures (a) and forces (b) versus separation. Theoretical
quantities are computed using the Drude model and tabulated
optical data (a) and plasma model (b). Solid lines and dashed line
show the confidence intervals with 95\% and 99.9\% confidence
levels, respectively.
In computations the values of the plasma frequency
$\omega_p=8.9\,$eV and the relaxation parameter
$\gamma=0.0357\,$eV were used, as determined in \cite{5b} for
Au films deposited on the test bodies from the resistivity
measurements.
}
\end{figure*}

\section{Drude model and the effect of finite plates}

Before considering the generalized plasma-like permittivity,
we briefly discuss the physical reasons why the Drude model
dielectric permittivity,
\begin{equation}
\varepsilon_D(\omega)=1-
\frac{\omega_p^2}{\omega(\omega+{\rm i}\gamma)},
\label{eq9}
\end{equation}
\noindent
fails to provide an adequate description of the thermal Casimir
force. The idea of this explanation belongs to Parsegian \cite{26}
who noticed that the Drude model is derived from the Maxwell
equations in an infinite metallic medium (semispace) with no
external sources, zero induced charge density and with nonzero
induced current $\mbox{\boldmath$j$}=\sigma_0\mbox{\boldmath$E$}$
($\sigma_0$ is the conductivity at a constant current).
In such a medium there are no walls limiting the flow of charges.
Physically the condition that the semispace is infinite means
that its extension is much longer than the extension of the
wave front (recently the role of finite size of the conductors
was also discussed in \cite{27a} in the case of two wires
interacting through the inductive coupling between Johnson
currents).

For real metal plates, however, the applicability conditions of
the Drude model are violated. The extension of the wavefront of
a plane wave is much longer than of any conceivable metal plate.
For plane waves of very low frequency, the electric field
$\mbox{\boldmath$E$}_i$ inside the plate is practically constant and it
is parallel to the boundary surface (see figure 4 where
{\boldmath$k$} and $\mbox{\boldmath$k$}_i$ are the wave vectors
outside and inside the plate, respectively).
Constant electric field $\mbox{\boldmath$E$}_i$ creates a
short-lived current of conduction electrons leading to the
formation of practically constant charge densities $\pm\rho$
on the opposite sides of the plate (see figure 4). As  a result,
both the electric field and the current inside the plate vanish.
The field outside the plate becomes equal to the superposition
of the incident field {\boldmath$E$} and the field
$\mbox{\boldmath$E$}_{\rho}$ produced by the charge densities
$\pm\rho$ \cite{27}. This process takes place in a very short time
interval of about $10^{-18}\,$s \cite{19}.
\begin{figure*}[t]
\vspace*{-14.4cm}
\includegraphics{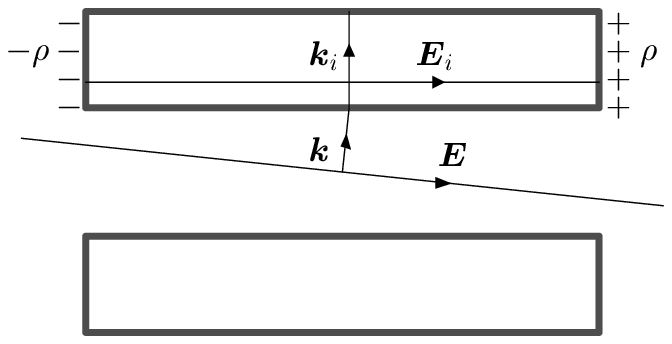}
\vspace*{-12.cm}
\caption{
The electromagnetic plane wave of a vanishing frequency
with a wave vector {\boldmath$k$} is incident on a metal
plate of finite size leading to the accumulation of
charges on its back sides.
}
\end{figure*}

One can conclude that a finite metal plate exposed to a plane wave of
very low frequency is characterized by zero current of conduction
electrons and nonzero induced current density. Thus, it cannot be
described by the Drude dielectric function (\ref{eq9}).
As to the plasma dielectric permittivity obtained from (\ref{eq9})
by putting $\gamma=0$, it leads to zero real current of conduction
electrons and admits only a displacement current. Because of this,
the plasma model does not allow the accumulation of charges on the
sides of a finite plate in the electromagnetic wave of low
frequency. Note also that for the plane waves of sufficiently high
frequency there is no problem in the application of the Drude
dielectric function. Thus, at $T=300\,$K the first Matsubara
frequency $\xi_1=2.47\times 10^{14}\,$rad/s and for plane waves
with $\omega\geq\xi_1$ the electric field $\mbox{\boldmath$E$}_i$
changes its direction so quickly that the average charge densities
on the plate sides are equal to zero, in accordance with the
applicability conditions of the Drude model.

\section{Generalized plasma-like permittivity}

The generalized plasma-like dielectric permittivity disregards
relaxation of conduction electrons (as does the usual plasma model)
but takes into account relaxation processes of core electrons.
It is given by
\begin{equation}
\varepsilon(\omega)=1-\frac{\omega_p^2}{\omega^2}+
\sum_{j=1}^{K}\frac{f_j}{\omega_j^2-\omega^2-{\rm i}g_j\omega},
\label{eq10}
\end{equation}
\noindent
where $\omega_j\neq 0$ are the resonant frequencies of core
electrons, $g_j$ are their relaxation parameters, and $f_j$ are
the oscillator strengths. The generalized plasma-like permittivity
was applied to describe the thermal Casimir force in \cite{18}.
As the usual plasma model, permittivity (\ref{eq10}) admits only
a displacement current and does not allow the accumulation
of charges on the
sides of a finite plate. It also leads to the same values (\ref{eq6})
of the reflection coefficients at zero frequency as the usual plasma
model. The values of parameters $f_j,\>\omega_j$ and $g_j$ can be found
by fitting the imaginary part of $\varepsilon(\omega)$ in (\ref{eq10})
to the tabulated optical data for the complex index of refraction.
For example, for Au the results of a 3-oscillator fit ($K=3$)
can be found in \cite{26}. Using the complete data in \cite{20},
the more exact 6-oscillator fit ($K=6$) for Au was performed in
\cite{5b}. The resulting values of the oscillator parameters are
presented in table 1.
\begin{table}[h]
\caption{The oscillator parameters for Au
found from the 6-oscillator fit to the tabulated optical
data.}
\label{tab:2}       
\begin{center}
\begin{tabular}{cccc}
\hline\noalign{\smallskip}
$j$ & $\omega_j\,$(eV) & $g_j\,$(eV) & $f_j\,(\mbox{eV}^2)$  \\
\noalign{\smallskip}\hline\noalign{\smallskip}
1 & 3.05 & 0.75 & 7.091 \\
2 & 4.15 & 1.85 & 41.46 \\
3 & 5.4{\ } & 1.0{\ } & 2.700 \\
4 & 8.5{\ } & 7.0{\ } & 154.7 \\
5 & 13.5 & 6.0{\ } & 44.55 \\
6 & 21.5 & 9.0{\ } & 309.6 \\
\noalign{\smallskip}\hline
\end{tabular}
\end{center}
\end{table}

Equation (\ref{eq10}) and table 1 were used together with the Lifshitz
formula (\ref{eq1}) to calculate the theoretical Casimir pressure in the
configuration of two parallel plates and Casimir force in the configuration
of a sphere above a plate. The results were compared with the
measurement data of \cite{5a,5b} and of \cite{16}, respectively.
The differences of the theoretical and mean experimental Casimir
pressures versus separation are shown in figure 5(a) as dots.
The differences of the theoretical and mean experimental Casimir
forces are shown as a function of separation in figure 5(b).
\begin{figure*}[t]
\vspace*{-12.4cm}
\includegraphics{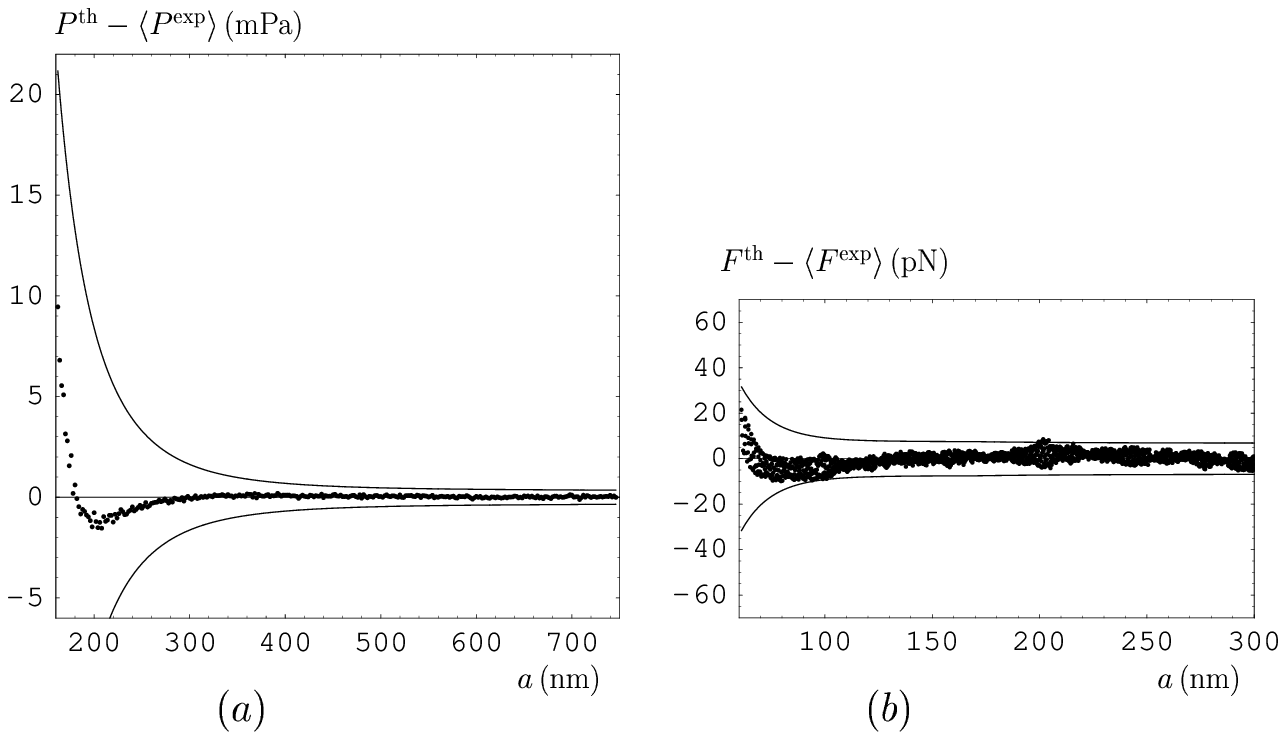}
\vspace*{-11.cm}
\caption{
Differences between the theoretical and experimental Casimir
pressures (a) and forces (b) versus separation. Theoretical
quantities are computed using the generalized plasma-like
permittivity (\ref{eq10}).
Solid lines
show the confidence intervals with 95\% confidence.
}
\end{figure*}
In both figures solid lines represent the borders of the 95\%
confidence intervals. As is seen in figures 5(a) and 5(b),
all dots are well inside the confidence intervals. Thus,
the generalized plasma-like dielectric permittivity (\ref{eq10})
combined with the Lifshitz formula is consistent with the
measurement data of both long- and short-separation measurements
of the Casimir force performed up to date.

\section{Kramers-Kronig relations and their generalizations}

An important advantage of the plasma-like dielectric permittivity (\ref{eq10})
is that it precisely satisfies the Kramers-Kronig relations. There is
some confusion in the literature concerning the Kramers-Kronig relations
in the case of usual plasma model which is characterized by entirely
real permittivity. In fact the form of Kramers-Kronig relations is
different depending on the analytic properties of the considered
dielectric permittivity. If
$\varepsilon(\omega)=\varepsilon^{\prime}(\omega)+{\rm i}
\varepsilon^{\prime\prime}(\omega)$ is regular at $\omega=0$,
the Kramers-Kronig relations take its simplest form \cite{27},
\begin{equation}
\varepsilon^{\prime}(\omega)=1+\frac{1}{\pi}{\rm P}
\int_{-\infty}^{\infty}
\frac{\varepsilon^{\prime\prime}(\xi)}{\xi-\omega}\,d\xi,
\qquad
\varepsilon^{\prime\prime}(\omega)=-\frac{1}{\pi}{\rm P}
\int_{-\infty}^{\infty}
\frac{\varepsilon^{\prime}(\xi)}{\xi-\omega}\,d\xi,
\label{eq11}
\end{equation}
\noindent
where the integrals are understood as a principal value.

However, if the dielectric permittivity has a simple pole
 at $\omega=0$, $\varepsilon(\omega)\approx 4\pi{\rm i}\sigma_0/\omega$,
the form of the Kramers-Kronig relations is different \cite{28},
\begin{equation}
\hspace*{-1cm}
\varepsilon^{\prime}(\omega)=1+\frac{1}{\pi}{\rm P}
\int_{-\infty}^{\infty}
\frac{\varepsilon^{\prime\prime}(\xi)}{\xi-\omega}\,d\xi,
\qquad
\varepsilon^{\prime\prime}(\omega)=-\frac{1}{\pi}{\rm P}
\int_{-\infty}^{\infty}
\frac{\varepsilon^{\prime}(\xi)}{\xi-\omega}\,d\xi+
\frac{4\pi\sigma_0}{\omega}.
\label{eq12}
\end{equation}
\noindent
In both cases of being regular and having a simple pole at $\omega=0$
dielectric permittivity the third dispersion relation, expressing
the dielectric permittivity along the imaginary frequency axis, is common:
\begin{equation}
\varepsilon({\rm i}\omega)=1+\frac{1}{\pi}{\rm P}
\int_{-\infty}^{\infty}
\frac{\xi\varepsilon^{\prime\prime}(\xi)}{\xi^2+\omega^2}\,d\xi.
\label{eq13}
\end{equation}

The standard derivation procedure \cite{27,28}, when applied to the
dielectric permittivities having a second-order pole at zero frequency
[i.e., with an asymptotic behavior
$\varepsilon(\omega)\approx -\omega_p^2/\omega^2$ when $\omega\to 0$]
leads to another form of Kramers-Kronig relations \cite{18},
\begin{equation}
\hspace*{-1.5cm}
\varepsilon^{\prime}(\omega)=1+\frac{1}{\pi}{\rm P}
\int_{-\infty}^{\infty}
\frac{\varepsilon^{\prime\prime}(\xi)}{\xi-\omega}\,d\xi
-\frac{\omega_p^2}{\omega^2},
\qquad
\varepsilon^{\prime\prime}(\omega)=-\frac{1}{\pi}{\rm P}
\int_{-\infty}^{\infty}
\frac{\varepsilon^{\prime}(\xi)+\frac{\omega_p^2}{\xi^2}}{\xi-\omega}\,d\xi.
\label{eq14}
\end{equation}
\noindent
In this case the third Kramers-Kronig relation (\ref{eq13}) also
is replaced with \cite{18}
\begin{equation}
\varepsilon({\rm i}\omega)=1+\frac{1}{\pi}{\rm P}
\int_{-\infty}^{\infty}
\frac{\xi\varepsilon^{\prime\prime}(\xi)}{\xi^2+\omega^2}\,d\xi
+\frac{\omega_p^2}{\omega^2},
\label{eq15}
\end{equation}
\noindent
i.e., it acquires an additional term.

It is easily seen that both the usual nondissipative plasma model
[given by (\ref{eq9}) with $\gamma=0$] and the generalized plasma-like
dielectric permittivity (\ref{eq10}) satisfy the Kramers-Kronig
relations (\ref{eq14}) and (\ref{eq15}) precisely.

\section{Thermodynamic test for the generalized plasma-like
permittivity}

As was noted in the Introduction, thermodynamics provides an important
test for the used model of dielectric properties. The Casimir entropy
computed by using the Lifshitz formula (\ref{eq1}) must vanish when the
temperature vanishes, i.e., the Nernst heat theorem must be satisfied.
This test was used to demonstrate the incompatibility of the Drude
model with the Lifshitz formula \cite{14,15}. The physical reasons
for this incompatibility were discussed in Section 3. In \cite{19}
the thermodynamic test was applied to the plasma-like dielectric
permittivity (\ref{eq10}). The low temperature behavior of the
Casimir entropy (\ref{eq7}) was found analytically under the conditions
\begin{equation}
T\ll T_{\rm eff}=\frac{\hbar c}{2ak_B}, \qquad
\alpha\equiv\frac{\lambda_p}{4\pi a}\ll 1,
\label{eq16}
\end{equation}
\noindent
where $\lambda_p=2\pi c/\omega_p$ is the plasma wavelength.
The Casimir entropy is given by
\begin{eqnarray}
&&
S(a,T)=\frac{3\zeta(3)k_B}{8\pi a^2}
\left(\frac{T}{T_{\rm eff}}\right)^2\left\{
\vphantom{\left(\frac{T}{T_{\rm eff}}\right)^2
\left[\left(\sum\limits_{j=1}^{K}C_j+2\right)
-\frac{\alpha^2}{\pi^2}\right]}
1+4\alpha \right.
\nonumber \\
&&\phantom{aaa}
-\frac{4\pi^3}{135\zeta(3)}\frac{T}{T_{\rm eff}}
\left(1+8\alpha+6\zeta(3)\alpha^3\sum\limits_{j=1}^{K}C_j\delta_j
-96\zeta(3)\alpha^4\sum\limits_{j=1}^{K}C_j\delta_j\right)
\nonumber \\
&&\phantom{aaa}
\left.
-\frac{40\zeta(5)}{3\zeta(3)}
\left(\frac{T}{T_{\rm eff}}\right)^2
\alpha^2\left[1+3\alpha
\left(\sum\limits_{j=1}^{K}C_j+2\right)
-12\alpha^2\right]\right\}.
\label{eq17}
\end{eqnarray}
\noindent
Here, the quantities $C_j$ and $\delta_j$ are expressed in terms of the
oscillator parameters
\begin{equation}
C_j=\frac{f_j}{\omega_j^2}, \qquad
\delta_j=\frac{cg_j}{2a\omega_j^2}.
\label{eq17a}
\end{equation}
Note that in \cite{19} the values of numerical coefficients in the
third lines of (38) and (41) are indicated incorrectly. Correct
values are obtained by the replacement of all $\pi^2$ with 1/12.
In the last term on the right-hand side of (33) in \cite{19}
$6/\pi^2$ should be replaced with $1/(2\pi^4)$.

The Casimir entropy defined in (\ref{eq17}) is nonnegative. It is seen that
\begin{equation}
S(a,T)\to 0 \quad \mbox{when} \quad T\to 0,
\label{eq18}
\end{equation}
\noindent
i.e., the Nernst heat theorem is satisfied. Thus, the plasma-like
dielectric permittivity is not only consistent with all experiments
performed up to date, but it also withstands the thermodynamic test.

\section{Metal-type semiconductors}

The above results are of importance not only for metals but also
for metal-type semiconductors. In more detail, the thermal Casimir
force between dielectrics and semiconductors is considered in \cite{29}.
Here, we briefly discuss only one point, i.e., how to account
for the influence of free charge carriers in metal-type semiconductors
where the density of these carriers is relatively high.

It is common \cite{20} to include the role of free charge carriers in
semiconductors by considering the dielectric permittivity of the form
\begin{equation}
\varepsilon(\omega)=\varepsilon_d(\omega)-
\frac{\omega_p^2}{\omega(\omega+{\rm i}\gamma)}.
\label{eq19}
\end{equation}
\noindent
Here, $\varepsilon_d(\omega)$ is the permittivity of high resistivity
(dielectric) semiconductor such that $\varepsilon_d(0)<\infty$.
This approach was used for the interpretation of most precise recent
experiment on the measurement of the Casimir force between metallic
sphere and semiconductor plate by means of an atomic force
microscope \cite{7}. In the measurement set under consideration the
density of charge carriers in a Si membrane was changed from
$5\times 10^{14}\,\mbox{cm}^{-3}$ to $2.1\times 10^{19}\,\mbox{cm}^{-3}$
through the absorption of photons from laser pulses.
In this differential experiment only the difference of the Casimir
forces, $\Delta F^{\rm exp}$, in the presence and in the absence of laser
pulse was measured. The experimental data on mean difference forces,
$\langle\Delta F^{\rm exp}\rangle$, was compared with the theoretical
difference forces, $\Delta F^{\rm th}$, computed using the Lifshitz theory.
\begin{figure*}[b]
\vspace*{-14.6cm}
\hspace*{-1.5cm}\includegraphics{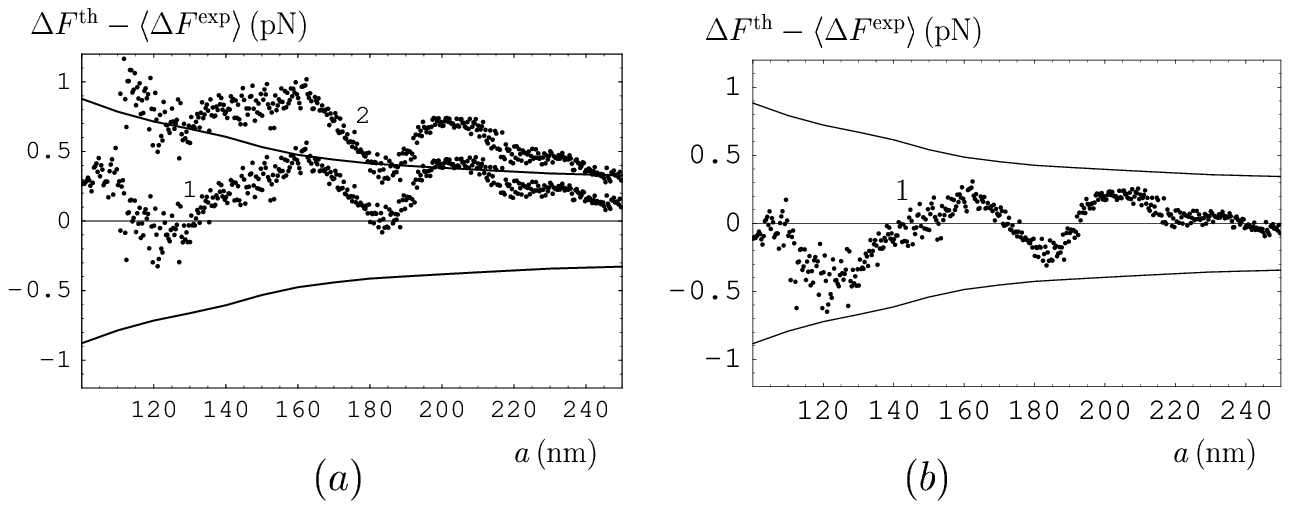}
\vspace*{-11.cm}
\caption{
Theoretical minus experimental
differences of the Casimir
forces versus separation.
In the absence of laser pulse the theoretical results for dots
labeled 1 are computed using $\varepsilon_d(\omega)$ and for dots
labeled 2 taking the dc conductivity of high resistivity
Si into account. In both cases $\varepsilon(\omega)$ from (\ref{eq19})
is used when the laser pulse is on (a). The same differences labeled
1 calculated using $\varepsilon_d(\omega)$ when the pulse is off and
$\tilde\varepsilon(\omega)$ from (\ref{eq20}) when the pulse is on
are shown in (b).
Solid lines show the confidence interval with 95\% confidence.
}
\end{figure*}
In figure 6(a) dots labeled 1 show the quantity
$\Delta F^{\rm th}-\langle\Delta F^{\rm exp}\rangle$
versus separation, where $\Delta F^{\rm th}$ is computed
under the assumption that in the absence of laser pulse high resistivity
Si is described by $\varepsilon_d(\omega)$, i.e., the effect of dc
conductivity is disregarded.  Dots labeled 2 show the same quantity, where
$\Delta F^{\rm th}$ is computed taking into account the dc conductivity
of high resistivity Si in the absence of laser pulse.
In both cases the dielectric permittivity (\ref{eq19})
with appropriate values of $\omega_p$ and $\gamma$ is used when
the laser pulse is on. Solid lines indicate the borders of 95\%
confidence intervals. As is seen in figure 6(a), the Lifshitz theory
taking the dc conductivity of high resistivity Si into account is
experimentally excluded. The physical explanation for this result
can be found in \cite{29}. It is notable also that, as was shown in
\cite{30}, the inclusion of the dc conductivity of a dielectric in
the Lifshitz theory results in the violation of the Nernst heat
theorem.

Bearing in mind that the Si plate has finite size, as was discussed in
Section 3, a question arises whether the use of the Drude-type
dielectric permittivity (\ref{eq19}) in the presence of laser pulses
for the calculation of the Casimir force is warranted.
To check this point we have recalculated the values of
$\Delta F^{\rm th}$ versus separation by using the dielectric
permittivity $\varepsilon_d(\omega)$ of high resistivity dielectric Si
in the absence of laser pulse and the plasma-like permittivity
\begin{equation}
\tilde\varepsilon(\omega)=\varepsilon_d(\omega)-
\frac{\omega_p^2}{\omega^2}
\label{eq20}
\end{equation}
\noindent
in the presence of pulse. The resulting quantity
$\Delta \tilde{F}^{\rm th}-\langle\Delta F^{\rm exp}\rangle$
is shown in figure 6(b) as dots labeled 1. As is seen from the
comparison of dots labeled 1 in figures 6(a) and 6(b), the use
of the generalized plasma-like permittivity (\ref{eq20}) leads to a bit
better agreement with data than the use of the Drude-type permittivity
(\ref{eq19}). However, it is not possible to give a statistically
meaningful preference to one of the models on these grounds because
in both cases most of the dots [of about 95\% in figure 6(a) and
100\% in figure 6(b)] are inside the confidence intervals.

Thus, although for metals (Au) we already have a decisive confirmation
of the fact that the generalized plasma-like permittivity is consistent with
experiments on measuring the Casimir force and that the Drude model is
excluded, for metal-type semiconductors such confirmation is still
lacking. It can be obtained in the proposed, more precise
experiments on
measuring the difference Casimir force between two sections of a
patterned Si plate of different dopant concentration \cite{31}.

\section{Conclusions and discussion}

To conclude, we have demonstrated that the use of the Drude dielectric
function in the Lifshitz formula is inconsistent with electrodynamics
in the case of finite plates. Instead, to calculate the thermal
Casimir force, one should use the generalized plasma-like permittivity
that disregards relaxation of free electrons but takes into account
relaxation due to interband transitions of core electrons.
This permittivity is the only one that is consistent with both
short- and long-separation measurements of the Casimir force at
$T=300\,$K. The generalized plasma-like permittivity satisfies
precisely the Kramers-Kronig relations. The use of this permittivity
also leads to a positive Casimir entropy which vanishes at zero
temperature in accordance with the Nernst heat theorem.
We have also demonstrated that the inclusion of dc conductivity of
high resistivity semiconductors is inconsistent with recent experiment
on the measurement of the difference Casimir force between metal sphere
and Si plate illuminated with laser pulses. We have presented two
theoretical descriptions for the dielectric properties of low
resistivity Si in the presence of laser pulse by means of the
Drude-type and plasma-type permittivities and concluded that available
experimental data is not of sufficient precision to discriminate
between them. This problem will be solved in the future using results
of the proposed experiment \cite{31}.

A more fundamental approach to the thermal Casimir force between
metals and metal-type semiconductors would require the consideration
of finite plates and a more sophisticated description of conduction
electrons that is far beyond the scope of the Lifshitz theory.

\section*{Acknowledgments}
The authors are grateful to G~Bimonte, R~S~Decca,
E~Fischbach, G~L~Klimchitskaya, D~E~Krause, K A Milton
and U~Mohideen for
helpful discussions.
VMM is
grateful to the Center of Theoretical Studies and Institute
for Theoretical Physics, Leipzig University for kind
hospitality.
This work  was supported by Deutsche Forschungsgemeinschaft,
Grant No.~436\,RUS\,113/789/0--3.

\end{document}